\documentclass[aps, prb, 10pt, preprint, superscriptaddress]{revtex4-1}

\usepackage{longtable}
\usepackage{array}
\usepackage{graphicx}
\usepackage{color}

\begin{document}
\title{Ferromagnetic ordering of linearly coordinated Co ions in LiSr$_2$[CoN$_2$]}

\author{T. J. Ball\'{e}}
 \affiliation{EP VI, Center for Electronic Correlations and Magnetism, Institute of Physics, University of Augsburg, D-86159 Augsburg, Germany}
\author{Z. Zangeneh}
 \affiliation{Institute for Theoretical Solid State Physics, IFW Dresden, Helmholtzstr. 20, D-01069 Dresden, Germany}
\author{L. Hozoi}
 \affiliation{Institute for Theoretical Solid State Physics, IFW Dresden, Helmholtzstr. 20, D-01069 Dresden, Germany} 
\author{A. Jesche}
\email[]{anton.jesche@physik.uni-augsburg.de}
 \affiliation{EP VI, Center for Electronic Correlations and Magnetism, Institute of Physics, University of Augsburg, D-86159 Augsburg, Germany}
\author{P. H\"ohn}
 \affiliation{Max-Planck-Institut f\"ur Chemische Physik fester Stoffe, N\"othnitzer Str. 40, D-01187 Dresden, Germany} 

\begin{abstract}
Li\textbf{Sr$_2$}[CoN$_2$] single crystals were successfully grown out of Li-rich flux. 
Temperature- and field-dependent measurements of the magnetization in the range of $T = 2 - 300$\,K and up to $\mu_{0}\textit{H} = 7$\,T as well as measurements of the heat capacity are presented.
Ferromagnetic ordering emerges below $T_C = 44$\,K and comparatively large coercivity fields of $\mu_0H = 0.3$\,T as well as pronounced anisotropy are observed upon cooling.
Polycrystalline samples of the Ca analog Li\textbf{Ca$_2$}[CoN$_2$] were obtained and investigated in a similar way. 
In both compounds Co manifests orbital contributions to the magnetic moment and large single-ion anisotropy that is caused by second-order Spin-orbit coupling.
Quantum chemistry calculations reveal a magnetic anisotropy energy of 7\,meV, twice as large as the values reported for similar Co \textit{d}$^{8}$ systems.
\end{abstract}

\maketitle

\section{Introduction}
The formation of local magnetic moments in solids has implications that range from the design of nanoscale data storage devices to the field of massive hard magnets\,\cite{Coey2002, Gutfleisch2011}.
Such magnetic materials usually require the presence of heavy transition-metal or rare-earth elements in order to provide a sufficiently large magnetic anisotropy. 
At the origin of the desired magnetic stability is an orbital contribution to the magnetic moment. 
In $3d$ transition-metal compounds, however, this contribution is normally almost fully suppressed by the crystal electric field.
Few exceptions are known that are not subject to such an 'orbital quenching', for example: LuFe$_{2}$O$_{4}$\,\cite{IIda1993, Wu2008}, NiO\,\cite{Kwon2000}, or molecular magnets with low coordination numbers\,\cite{Zadrozny2012, Power2012}.
More recently, orbital moments, which give rise to large magnetic anistropy, were found in the comparatively simple, inorganic materials Li$_2$(Li$_{1-x}T_{x}$)N with $T$ = Fe\,\cite{Jesche2014b} and $T = $ Mn, Co, Ni\,\cite{Jesche2015} as well as Li$_2$Sr[Li$_{1-x}$Fe$_x$N]$_2$\,\cite{Hohn2016}.
Similar properties are observed in linear Fe(I) \cite{Zadrozny2013} and Co(I) \cite{Meng2015} complexes.
The unusual presence of orbital moments in such materials is related to a rare structural motif: a linear coordination of the magnetic ion \cite{Zadrozny2013}. 
As such, the 'effective molecule' or chain is not subject to a Jahn-Teller distortion\,\cite{Jahn1937}.

\begin{figure}
 \begin{center}
  \includegraphics[width=0.6\textwidth]{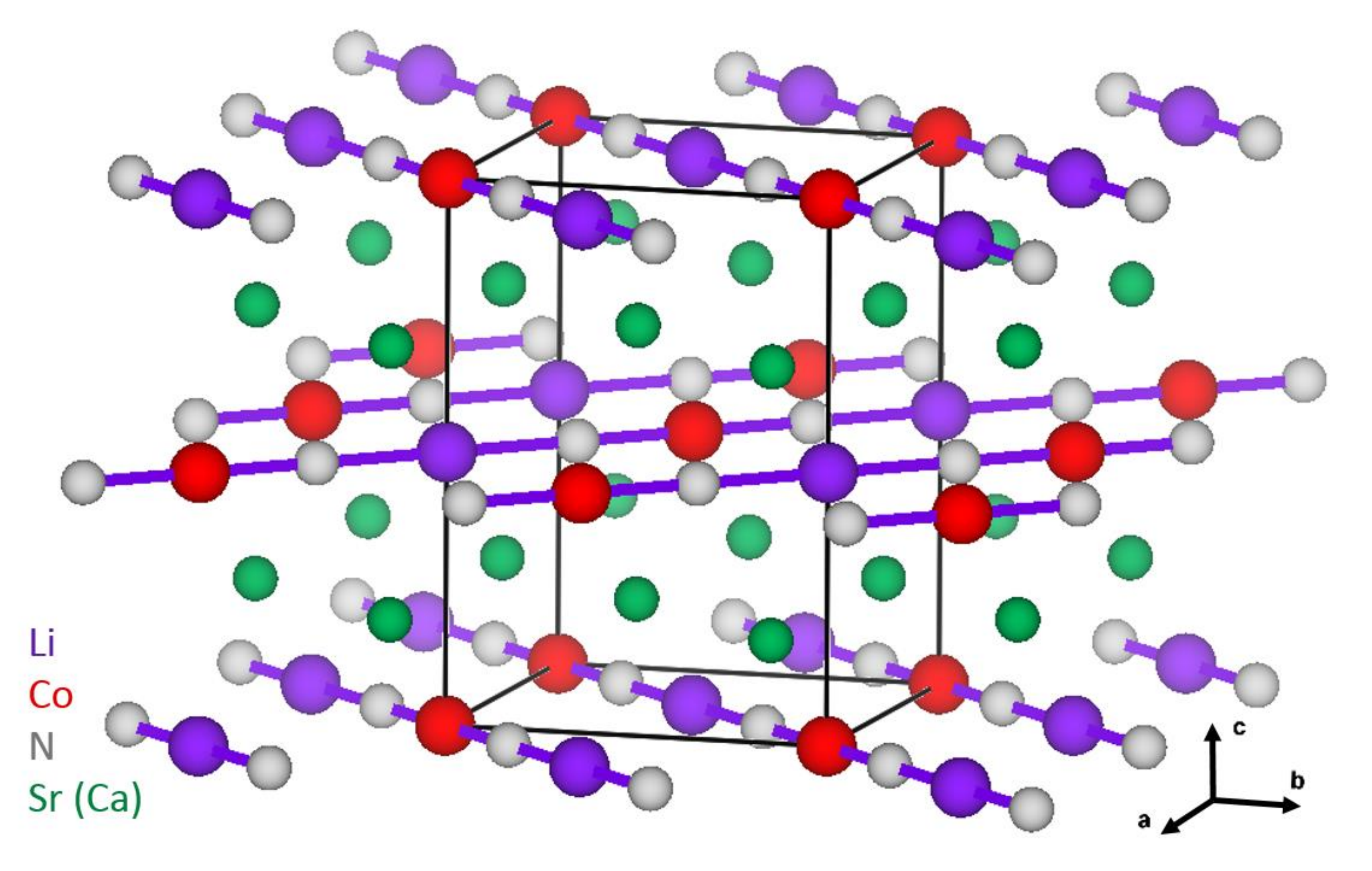}
  \caption{Crystal structure of Li\textit{AE}$_2$[CoN$_2$] (\textit{AE} = Sr, Ca). The effective N-Co-N linear molecules form the basic magnetic unit and are oriented along [$1\,1\,0$] and [$1\,\bar{1}\,0$].}
	\label{crystalStructure}
 \end{center}
\end{figure}

Here we present another material that contains an effective linear molecule: Li\textbf{Sr$_2$}[CoN$_2$] with $\cdot\cdot$N-Co-N-Li-N$\cdot \cdot$ chains running along $<$1\,1\,0$>$.
The crystal structure is shown in Fig.\,\ref{crystalStructure}. 
Li\textbf{Sr$_2$}[CoN$_2$] crystallizes in a tetragonal lattice with the space group \textit{P}4$_{2}$/\textit{mnm}. The structure is built from linear anions [CoN$_{2}$]$^{5-}$ that are oriented perpendicular to the \textit{c}-axis. 
The resulting $\cdot\cdot$N-Co-N-Li-N$\cdot\cdot$ chains run along [$1\,1\,0$] and [$1\,\bar{1}\,0$] in an alternating fashion and form layers that are separated by Sr ions. 

Li\textbf{Sr$_2$}[CoN$_2$] belongs to a series of phases with ideal composition Li\textit{AE}$_{2}$[\textit{M}N$_{2}$] (\textit{AE} = Ca, \textit{M} = Fe \cite{Klatyk1999c}; \textit{AE} = Sr, \textit{M} = Co \cite{Hohn1992}; \textit{AE} = Ca, \textit{M} = Cu \cite{Jager1992}) isotypic to $\alpha$-Li$_{3}$[BN$_{2}$] \cite{Cenzual1991, Yamane1987}, in which Li/\textit{M} substitution may be observed on both the \textit{M} \cite{Klatyk1999c, Jager1992} as well as the Li position \cite{Jager1992}. Besides Li/\textit{M} substitution, also partial occupation of the Li position is discussed.  For \textit{M} = Ni \cite{Gudat1992}, no ordering Li/Ni is observed, resulting in a halved unit cell Ca[Ni$_{1-x}$Li$_{x}$N] isotypic to YCoC \cite{Gerss1986}. 

The structure presents well isolated magnetic centers in a truly linear, two-fold coordination, that manifests in a large ratio between the distance of the nearest-neighbors of Co to the next-nearest-neighbors of Co of approximately 1.8. 
Single crystals large enough to investigate the magnetic anisotropy of Li\textbf{Sr$_2$}[CoN$_2$] were grown and a thorough magnetic and structural characterization is presented, complemented by heat capacity measurements and quantum chemistry calculations. 
For the isoelectronic Li\textbf{Ca$_2$}[CoN$_2$], polycrystalline samples were obtained and characterized.

\section{Methods}
In view of the sensitivity of the starting materials and of the reaction products to moisture and air, all manipulations associated with sample preparation and handling were performed in inert atmosphere using an Ar-filled (Praxair, $>$ 99.999\,\%, purified with AirLiquide Oxisorb catalyst) glove box (MBraun, \textit{p}(O$_{2}$)/\textit{p}$_{0}$ $<$ 0.1 ppm, \textit{p}(H$_{2}$O)/\textit{p}$_{0}$ $<$ 0.1 ppm). 

Laboratory powder X-ray diffraction data of selected and finely ground black crystals of both Li\textbf{Sr$_2$}[CoN$_2$] and Li\textbf{Ca$_2$}[CoN$_2$] were collected on a Huber G670 imaging plate Guinier camera (2$\Theta_{max}$ = 100$^\circ$) using a curved germanium (111) monochromator and Cu-\textit{K}$_{\alpha1}$ radiation at 293(1) K. The powder samples were placed between Kapton foils to avoid degradation in air. Preliminary data processing was done using the WinXPow program package \cite{XPow2003}.

X-ray diffraction intensity data of single crystals of both phases, Li\textbf{Sr$_2$}[CoN$_2$] and Li\textbf{Ca$_2$}[CoN$_2$], sealed in glass capillaries were collected at room temperature on a Rigaku AFC7 diffractometer equipped with a Saturn 724+ CCD area detector (Mo-\textit{K}$_{\alpha}$ radiation). 

Further details on the crystal structure investigations may be obtained from the Fachinformationszentrum Karlsruhe, 76344 Eggenstein-Leopoldshafen, Germany (fax: (+49)7247-808-666; email: crysdata@fiz-karlsruhe.de,\\ http://www.fiz-karlsruhe.de/request$\_$for$\_$deposited$\_$data.html) on quoting the deposition numbers CCDC 1875937 and CCDC 1875936 (re-evaluated) for Li\textbf{Sr$_2$}[CoN$_2$] and CCDC 1875935 for Li\textbf{Ca$_2$}[CoN$_2$], respectively.

To determine the orientation of the Li\textbf{Sr$_2$}[CoN$_2$] sample, Laue-back-reflection-patterns were taken with a digital Dual FDI NTX camera manufactured by Photonic Science (tungsten anode, U\,=\,20\,kV).
Chemical analysis was performed using inductively coupled plasma optical emission spectroscopy (ICP-OES, Vista-MPX). To this end, the samples were dissolved in a mixture of deionized water and dilute hydrochloric acid (37\,\%) in a ratio of 23:2.
The magnetic measurements were performed using a 7\,T Magnetic Property Measurement System (MPMS3 manufactured by Quantum Design), the heat capacity was measured in a 14\,T Physical Properties Measurement System (PPMS manufactured by Quantum Design). 

{\it Ab initio} calculations were carried out using the quantum chemistry package {\sc molpro} \cite{Werner2012}.
We applied all-electron Douglas-Kroll basis sets of triple-zeta quality for the Co ion \cite{Balabanov05} 
and all-electron triple-zeta basis functions for the two nearest-neighbor nitrogen ligands \cite{Dunning1989} and the Li atoms \cite{Prascher2011}.
Large-core pseudopotentials \cite{Fuentealba1985} were used for modeling the Sr$^{2+}$ (Ca$^{2+}$) species, incorporating all occupied shells. The solid-state surroundings were modelled as a large array of point charges fitted to reproduce the crystal Madelung field in the cluster region \cite{Klintenberg2000}.

\section{Crystal growth and basic structural characterization} \label{Crystal growth}

\begin{figure}
 \begin{center}
  \includegraphics[width=0.6\textwidth]{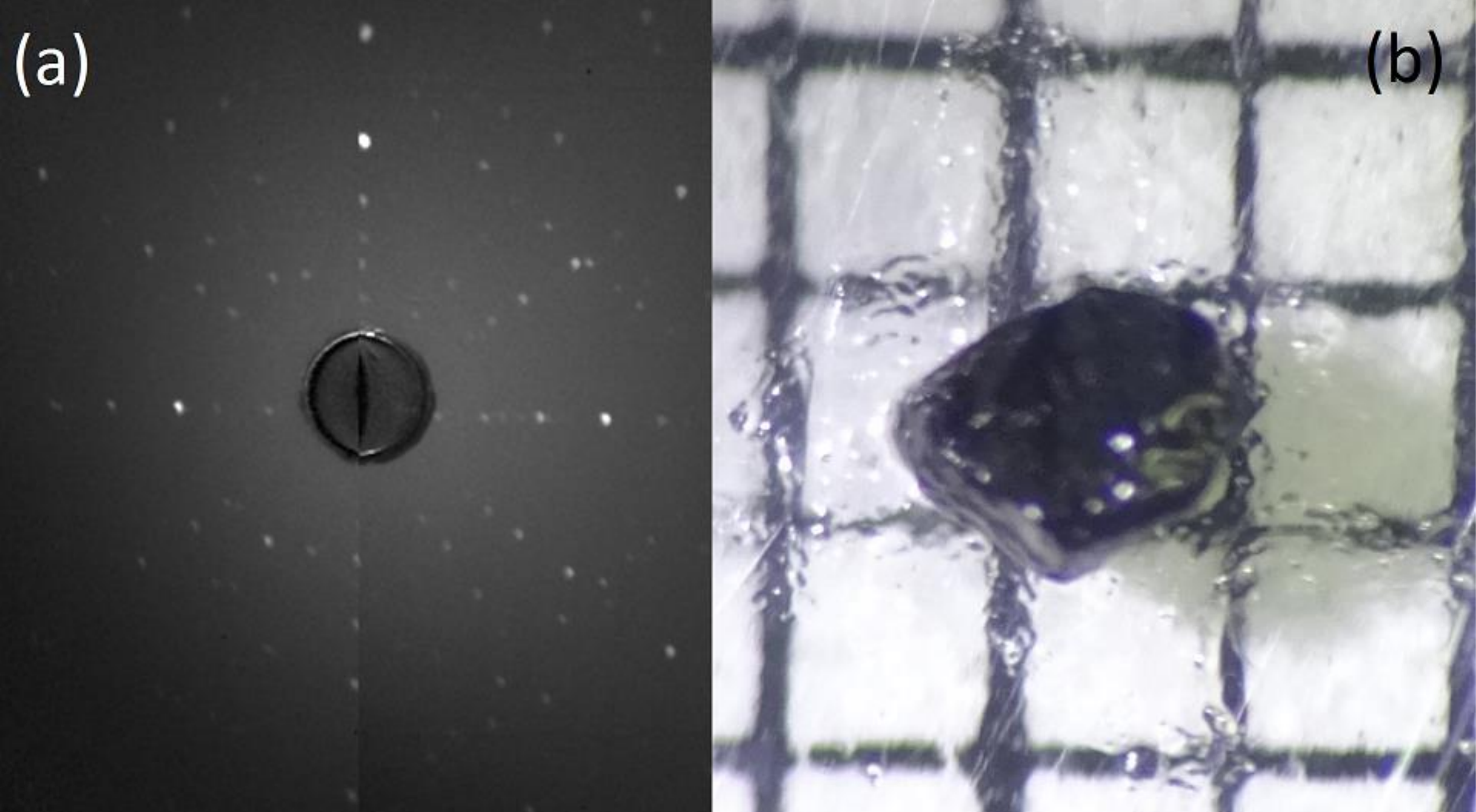}
	\caption{(\textbf{a}) Laue-back-reflection-pattern of Li\textbf{Sr$_2$}[CoN$_2$]. The beam was oriented parallel to the crystallographic \textit{c}-axis and the fourfold symmetry along the tetragonal \textit{c}-axis is clearly visible. (\textbf{b}) Photo of the corresponding single crystal on a millimeter grid. The sample was embedded in vacuum grease in order to prevent oxidation.}
	\label{Laue}
 \end{center}
\end{figure}

Li$_{3}$N, Ca$_{3}$N$_{2}$, and Sr$_{2}$N were synthesized from the metallic elements (Li: Evochem, 99.9\,\%; Ca: Alfa Aesar, distilled dendritic pieces, 99.98\,\%; Sr: Alfa Aesar, distilled dendritic pieces, 99.95\,\%) and N$_{2}$ (Praxair, 99.999\,\%, further purified by molecular sieve, 3\,\AA, Merck), the metallic elements were used without further purification.
Sample preparation was carried out employing a modified high temperature centrifugation-aided filtration (HTCAF) technique \cite{Hohn2016}. The materials were filled in a Ta ampoule equipped with a strainer, welded shut, and subsequently sealed in a quartz tube \cite{Canfield2001, Jesche2014c}. The quartz tube was placed in a stainless steel container fitted with quartz wool and put in a box furnace. The temperature program involved heating to 1023\,K by 100 Kh$^{-1}$, annealing for 2\,h and subsequent cooling with 2 Kh$^{-1}$ to 723\,K for the Ca and 823\,K for the Sr phase, respectively. After turning the stainless steel container by 180$^\circ$, the sample was centrifuged at 3000 min$^{-1}$, removing the flux consisting mostly of lithium and further unspecified impurities. 

Black agglomerates of tetragonal Li\textbf{Ca$_2$}[CoN$_2$] crystals up to several mm in size of were grown from a mixture of 0.078\,g Ca$_{3}$N$_{2}$, 0.094\,g Co (Alfa Aesar, 99.998\,\%) and 0.257\,g Li corresponding to a molar ratio Li:Ca:Co:N\,=\,23:1:1:0.7.

Large single crystals of Li\textbf{Sr$_2$}[CoN$_2$] up to several mm in size and used for the magnetic investigations were grown from a mixture of 0.514\,g Li, 0.106\,g Li$_{3}$N, 1.600\,g Sr$_{2}$N, and 0.127\,g Co giving rise to a molar ratio Li:Sr:Co:N\,=\,34:8:4:5. 
Both, black hexagonal (Li$_{3}$N) and tetragonal Li\textbf{Sr$_2$}[CoN$_2$] crystals were observed on the bottom of the ampoule after opening the Ta tube in the glove box. 
Powder X-ray diffraction revealed lattice parameters of \textit{a} = 541.73(3) pm and \textit{c} = 731.23(5) pm.

The chemical composition of two Li\textbf{Sr}$_{2}$[Co$_{x}$N$_{2}$] single crystals was determined from the ratio of Li, Co and Sr by ICP-OES, assuming a fully occupied Sr site. The two samples show Sr:Co-ratios of 2:1.02 and 2:1.01 as well as Sr:Li-ratios of 2:0.81 and 2:0.84, respectively, indicating significant Li-vacancies.

For the iso-structural Li\textbf{Ca$_2$}[CoN$_2$], three polycrystalline samples were measured and gave Ca:Co-ratios of 2:0.93, 2:0.94 and 2:0.96, respectively.
Those measurements revealed a considerable Li-excess of roughly three Li per formula unit that is caused by Li-rich flux remnants hidden in the polycrystalline material. 
Therefore, only the ratio Ca:Co was used to determine the amount of Co and a fully occupied Li-site (2\textit{b} Wyckoff site) is assumed. 

Furthermore, a Laue-back-reflection-pattern of Li\textbf{Sr$_2$}[CoN$_2$] was recorded in order to verify the orientation of the large single crystals (Fig.\,\ref{Laue}\textbf{a}). 
The incident X-ray-beam was oriented parallel to the $c$-axis along the surface normal of the largest facet of the crystal. 
The fourfold-symmetry of the tetragonal structure is reflected in the measured pattern. 
Fig.\,\ref{Laue}\textbf{b} shows a single crystal of Li\textbf{Sr$_2$}[CoN$_2$] on a millimeter grid.

\section{Single crystal X-ray diffraction}

\begin{table}[htp]
\begin{center}
\caption{Selected crystal structure data and results obtained from single crystal X-ray diffraction of Li\textit{AE}$_2$[CoN$_2$] (\textit{AE} = Ca, Sr).}

\begin{tabular}{l|l|l|l}
\hline
\hline

Compound  & Li\textbf{Ca$_2$}[CoN$_2$]  &  Li\textbf{Sr$_2$}[CoN$_2$] & Li\textbf{Sr$_2$}[CoN$_2$] \\
  &  & & re-evaluated \\
 \hline
Space group & \textit{P}4$_{2}$/\textit{mnm} & \textit{P}4$_{2}$/\textit{mnm} & \textit{P}4$_{2}$/\textit{mnm} \\

\textit{a} [pm] & 527.87(6) & 543.30(17) & 542.4(1) \\

\textit{c} [pm] & 672.91(11) & 735.3(4) & 730.7(2)\\

\textit{V} [10$^{6}$pm$^{3}$] & 187.50(5) & 217.04(18) & 214.97(10)\\

\textit{Z} & 2 & 2 & 2 \\

$\lambda$ [\AA] & Mo-\textit{K}$_{\alpha}$ & Mo-\textit{K}$_{\alpha}$ & Mo-\textit{K}$_{\alpha}$ \\

\textit{T} [K] & 293(1) & 293(1) & 293(1) \\

2$\Theta$ Range [$^\circ$] & 9.816-59.990 & 9.330-59.75 & 9.360-55.084\\

$\rho$ [g cm$^{-3}$] & 3.083 & 4.118 & 4.158 \\

\textit{F}(000) & 168 & 240 & 240 \\

Refl. meas./uni. & 2223/169 & 2659/192 & 265/151 \\

Ref. parameters & 16 & 16 & 16\\

$\mu$ [mm$^{-1}$] & 7.073 & 28.060 & 28.331 \\

$\frac{R_{1} (F_{0} > 4\sigma(F_{0}))}{R_{1} (all)}$ & 0.024 / 0.025 & 0.029 / 0.036 & 0.035 / 0.047 \\

w\textit{R}$_{2}$ / GooF & 0.048 / 1.362 & 0.069 / 1.149 & 0.098 / 1.120 \\

Res.$^{ }e^{-}$ [10$^{-6 }$pm$^{-3}$] & 0.463, -0.309 & 0.992, -1.388 & 0.862, -1.027\\

Programs & SHELXL* ** & SHELXL* ** & SHELXL* ** \\
& 	DIAMOND***  & 	DIAMOND*** & DIAMOND*** \\
\hline
\hline
\end{tabular}
\end{center}
  *\cite{Sheldrick2008} **\cite{Sheldrick2015} ***\cite{Putz2017}
\label{Structure Data}
\end{table}

Smaller single crystals for the crystal structure determination of Li\textbf{Sr$_2$}[CoN$_2$] were grown from a mixture of 0.251\,g Li, 0.130\,g Li$_{3}$N, 0.361\,g Sr$_{2}$N, and 0.037\,g Co corresponding to the molar ratio Li:Sr:Co:N\,=\,76:9:1:6. The temperature program involved heating to 923\,K by 100 Kh$^{-1}$, annealing for 2\,h and subsequent cooling with 1 Kh$^{-1}$ to 573\,K. Li\textbf{Sr$_2$}[CoN$_2$] was obtained as a side phase only. 
Given that structural parameters (see below) of those smaller single crystals are similar to the ones obtained for the larger single crystals (see previous section), we assume that there are no major differences (except for sample size) caused by the different temperature profile and composition of the starting materials.

\begin{table}[htp]
\begin{center}
\caption{Wyckoff positions, fractional atomic coordinates and isotropic displacement parameters for Li\textit{AE}$_2$[CoN$_2$]. The three values shown for \textit{x}, \textit{occ.} and \textit{U}$_{eq}$ correspond to \textit{AE} = Ca, Sr, Sr re-evaluated, respectively. See text.}

\begin{tabular}{l|l|l|l|l|l|l}
\hline \hline
Atom  & Site & \textit{x} & \textit{y} & \textit{z} & \textit{occ.} & \textit{U}$_{eq}$ [10$^{4}$ pm$^{2}$]\\
 \hline
 
\textit{AE} & 4\textit{d} & 0 & 1/2 & 1/4 & 1 & 0.0110(3) \\
&&&&&&0.0115(5) \\
&&&&&&0.0133(9) \\

Co1 & 2\textit{a} & 0 & 0 & 0 & 0.984(5)  & 0.0097(3)  \\
&&&&&0.994(11)  &0.0118(7) \\
&&&&&0.95(2) &0.0112(11) \\

Li1 & 2\textit{a} & 0 & 0 & 0 & 0.016  & 0.0097(3)  \\
&&&&&0.006 &0.0118(7) \\
&&&&&0.05 &0.0112(11)\\

N & 4\textit{g} & 0.2401(4) & \textit{x} & 0 & 1 & 0.0129(6) \\
&& 0.2335(5) &&&&0.0169(12) \\
&& 0.2370(11)&&&& 0.015(2) \\

Li2 & 2\textit{b} & 0 & 0 & 1/2 & 0.994(6)  & 0.019(3) \\
&&&&&0.974(17) &0.014(7)\\
&&&&& 0.97(3)&  0.034(14)\\

Co2 & 2\textit{b} & 0 & 0 & 1/2 & 0.006 & 0.019(3)  \\
&&&&&0.026 & 0.014(7)\\
&&&&& 0.03 &0.034(14)\\
\hline
\hline

\end{tabular}
\label{Wyckoff positions}
\end{center}
\end{table}

%The comparison of data obtained from single crystal X-ray diffraction and Powder X-ray diffraction (see Sec. \ref{Crystal growth}) yielded similar lattice parameters. Both are in good agreement with the chemical analysis, revealing a fully occupied Co-site.

The reflections of the X-ray powder diagrams of both phases were indexed in the tetragonal system. 
Indexing of the single-crystal diffraction images yielded tetragonal unit cells of Laue class 4/\textit{mmm}.
Systematic extinctions were identified leading to space group, \textit{P}4$_{2}$/\textit{mnm} (\#136) and its tetragonal subgroups \textit{P}4$_{2}$\textit{nm} (\#102) and \textit{P}$\bar 4$\textit{n}2 (\#118); all further calculations were executed in the highest-symmetry space group. 
 
%with the lattice parameters \textit{a} = 543.30(17) pm, \textit{c} = 735.3(4) pm and \textit{a} = 527.87(6) pm, \textit{c} = 672.91(11) pm, respectively. 
Literature data \cite{Hohn1992} were employed for the initial structural model \cite{FootnotePeter}, the subsequent refinement was done with the SHELXL-2018/1 software \cite{Sheldrick2008, Sheldrick2015}. After several cycles of least squares refinement and difference Fourier mapping, all atomic positions were found and all positions were refined anisotropically.
Furthermore, the original single crystal diffraction data of the 1992 publication on Li\textbf{Sr$_2$}[CoN$_2$] \cite{Hohn1992} was also re-evaluated\,\cite{FootnotePeter}. These crystals were prepared by annealing mixtures of Li, Sr$_{2}$N and Co with the molar ratio 6:3:1 in nitrogen atmosphere at ambient pressure and 1123\,K for 48 hours and subsequent cooling to room temperature with 100 Kh$^{-1}$.

\begin{table}[htp]
\begin{center}
\caption{Selected interatomic distances in Li\textit{AE}$_2$[CoN$_2$].}

\begin{tabular}{l|c|l|l|l}
\hline
\hline
Distance  & Occurrence & \textit{AE} = Ca & \textit{AE} = Sr & \textit{AE} = Sr\\
 & per layer & & & re-evaluated \\
 \hline
Co - N & 2$\times$ & 179.3(3) & 179.4(4) & 181.8(9) \\
Co - \textit{AE} & 4$\times$ & 312.99(3) & 328.00(9) & 326.99(6) \\
\textit{AE} - N & 4$\times$ & 251.37(3) & 266.18(8) & 265.04(5) \\
Li - N & 2$\times$ & 194.0(3) & 204.8(4) & 201.7(9) \\
Co - Co ($\|$ and $\bot$ chain,& 4$\times$ & 746.52(6) & 768.34(17) & 767.07(10) \\
within the \textit{ab}-plane) & & & & \\
Co - Co (along \textit{a}, \textit{b}) & 4$\times$ & 527.87(6) & 543.30(17) & 542.40(10) \\
Co - Co (2 layers)& 8$\times$ & 502.52(4) & 531.75(15) & 529.70(8) \\
Co - Co (along \textit{c}) & 2$\times$ & 672.91(11) & 735.3(4) & 730.7(2) \\

\hline
\hline
 
\end{tabular}
\label{Distances}
\end{center}
\end{table}

Crystallographic data, atomic coordinates, and isotropic thermal displacement parameters from structure refinements obtained by single crystal X-ray diffraction data are summarized in Tables I and \ref{Wyckoff positions}, relevant interatomic distances are listed in Table \ref{Distances}. The refinements of the occupation parameters of all positions resulted in no significant deviations from unity in all cases. No significant deficiencies were observed in any of the crystal structures of Li\textbf{Sr$_2$}[CoN$_2$] and Li\textbf{Ca$_2$}[CoN$_2$]; the amount of substitution Co/Li observed on both the Co and Li positions is always below 5(2)\,\%.
In other words, one of the positions between N neighbors (2\textit{a} Wyckoff site) is almost completely occupied by Co whereas the neighboring one (2\textit{b} Wyckoff site) is merely, if at all, substituted. 

%\newpage
\section{Magnetic Properties}

\begin{figure}
 \begin{center}
  \includegraphics[width=0.6\textwidth]{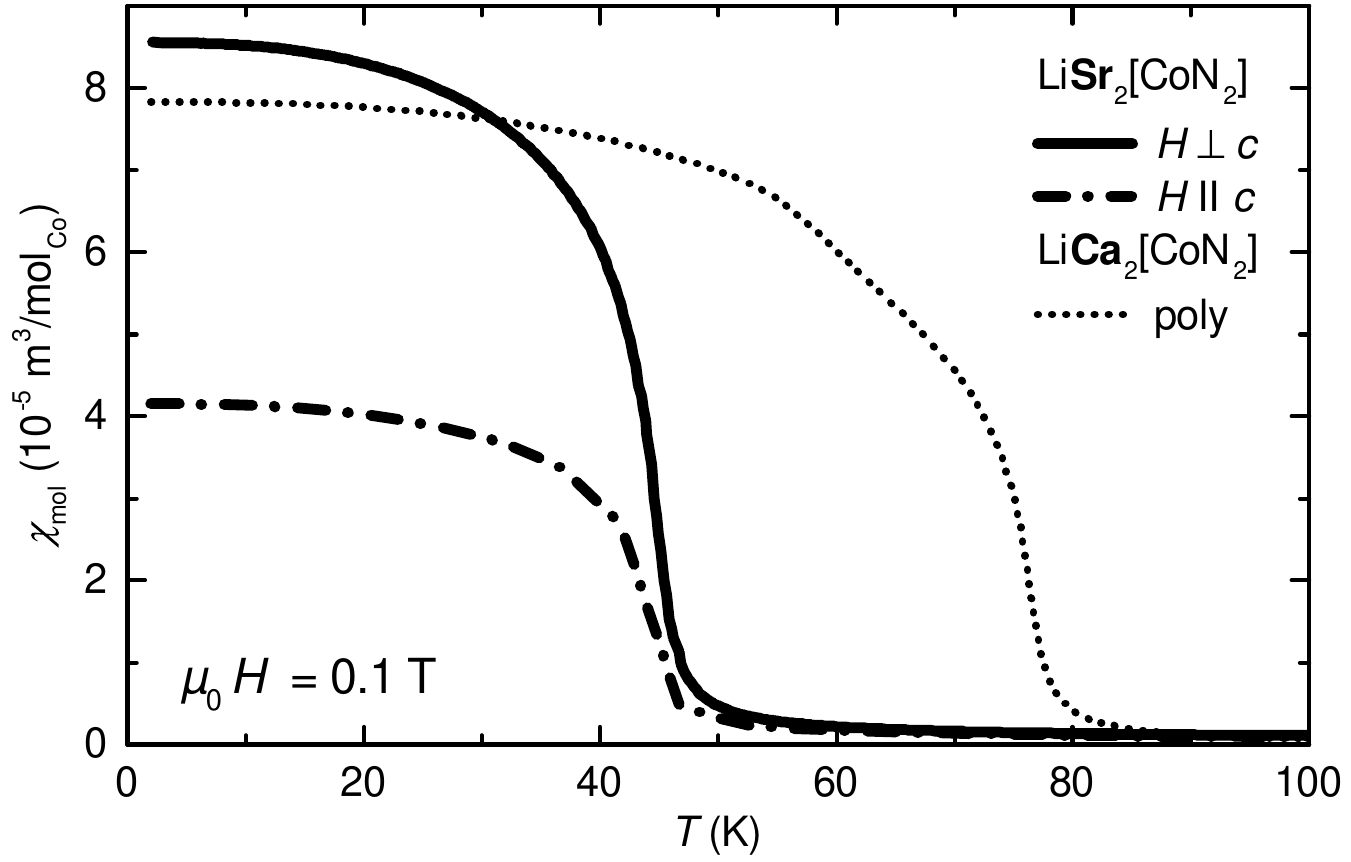}
	\caption{Magnetic susceptibility $\chi$\,=\,\textit{M}/\textit{H} per mol Co as a function of temperature for an applied magnetic field of $\mu_{0}\textit{H}$\,=\,0.1 T.
The solid and dashed lines show the measurements on plate-like Li\textbf{Sr$_2$}[CoN$_2$] with field applied perpendicular and parallel to the crystallographic \textit{c}-axis. Ferromagnetic ordering is observed at $T_{\mathrm{C}}$\,=\,44\,K. The dotted line shows the measurement on polycristalline Li\textbf{Ca$_2$}[CoN$_2$], that orders ferromagnetically at $T_{\mathrm{C}}$\,=\,76\,K.}
	\label{MvT}
 \end{center}
\end{figure}

\begin{figure}
 \begin{center}
  \includegraphics[width=0.6\textwidth]{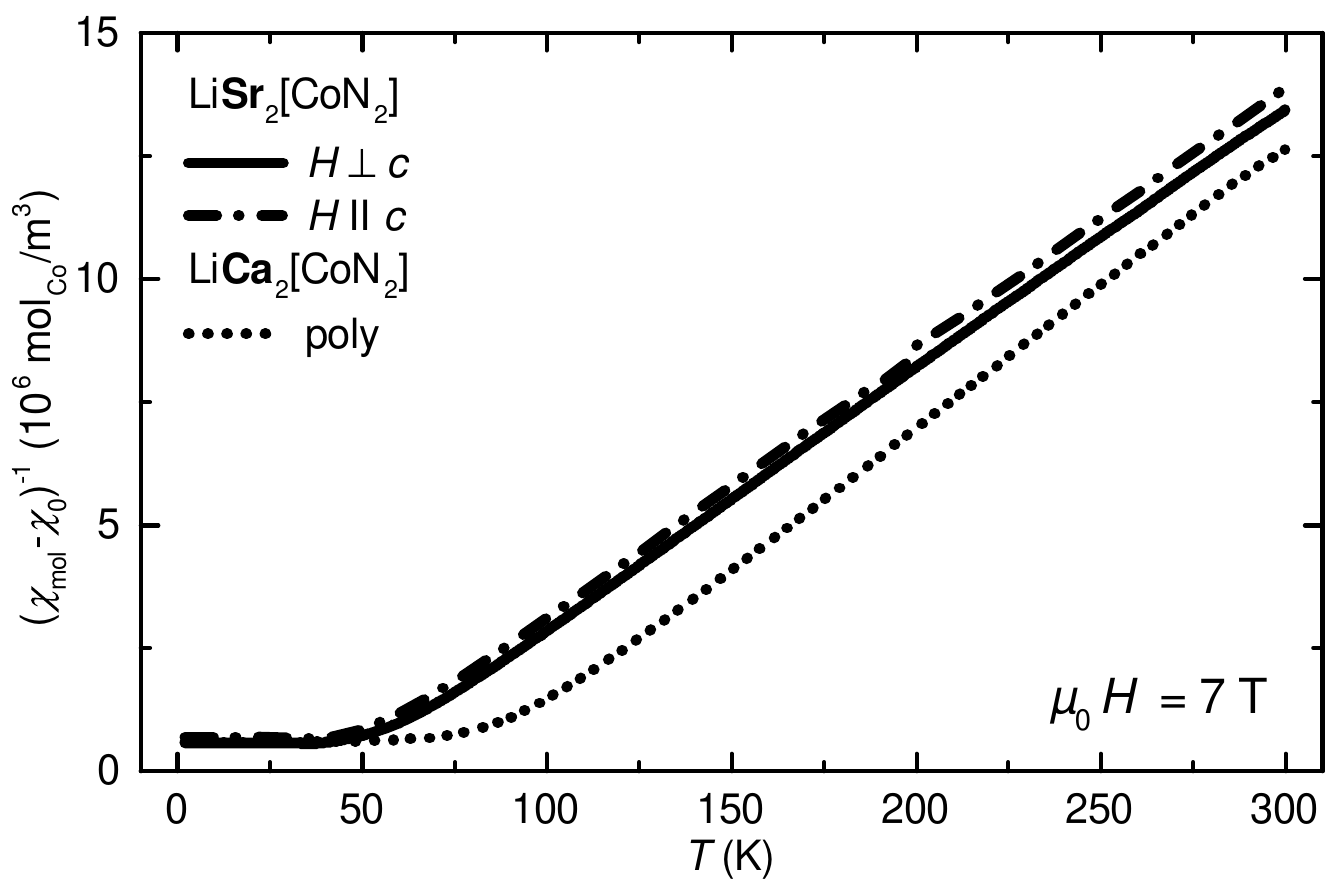}
	\caption{Inverse magnetic susceptibility, $1/(\chi_{\mathrm{mol}}-\chi_{0})$ (\textit{T}), per mol Co as a function of temperature for an applied magnetic field of $\mu_{0}\textit{H}$\,=\,7\,T. The extracted Weiss-temperatures of $\Theta_{\mathrm{W}}$\,=\,46\,K and $\Theta_{\mathrm{W}}$\,=\,42\,K for Li\textbf{Sr$_2$}[CoN$_2$] with \textit{H}\,$\perp$\,\textit{c} and \textit{H}\,$\parallel$\,\textit{c}, respectively, are in good agreement with the \textit{T}$_{\mathrm{C}}$ inferred from $\chi(T)$ in small applied fields. For polycristalline Li\textbf{Ca$_2$}[CoN$_2$] $\Theta_{\mathrm{W}}$ amounts to 78\,K.}
	\label{MvT Vgl}
 \end{center}
\end{figure}

Figure\,\ref{MvT} shows the temperature dependent magnetic susceptibility, $\chi(T)$, per mol Co for an applied field of $ \mu_{0}\textit{H}$\,=\,0.1\,T. 
The measurements on Li\textbf{Sr$_2$}[CoN$_2$] were performed with both \textit{H}\,$\perp$\,\textit{c} and \textit{H}\,$\parallel$\,\textit{c}.
%Temperature dependence was investigated in the range between 2\,K and 300\,K. 
The strong increase of $\chi(T)$ upon cooling indicates ferromagnetic ordering. 
A Curie temperature of $T_{\mathrm{C}}$\,=\,44\,K is inferred from the minimum in the derivitive $\frac{\mathrm{d}\chi}{\mathrm{d}T}$. 
For polycrystalline Li\textbf{Ca$_2$}[CoN$_2$] a Curie temperature of $T_{\mathrm{C}}$\,=\,76\,K was determined.

Figure\,\ref{MvT Vgl} shows the inverse magnetic susceptibility, $1/(\chi_{\mathrm{mol}}-\chi_{0})$ (\textit{T}), in a magnetic field of $\mu_{0}\textit{H}$\,=\,7\,T. 
Li\textbf{Sr$_2$}[CoN$_2$] follows a Curie-Weiss-law above $T \sim75$\,K with Weiss-temperatures of $\Theta_{\mathrm{W}}$\,=\,46\,K and $\Theta_{\mathrm{W}}$\,=\,42\,K for \textit{H}\,$\perp$\,\textit{c} and \textit{H}\,$\parallel$\,\textit{c}, respectively, in good agreement with the transition temperatures determined from $\chi(T)$. 
Small, but significant, deviations from Curie-Weiss-behavior appear for \textit{H}\,$\perp$\,\textit{c}. 
This is reflected in the sizable, temperature-independent $\chi_{0} = 2.0 \cdot 10^{-8}$\,\,m$^3$\,mol$_{\rm Co}$ that was also inferred from the Curie-Weiss fit.
For \textit{H}\,$\parallel$\,\textit{c}, on the other hand, a smaller $\chi_{0} = -9.6 \cdot 10^{-10}$\,m$^3$\,mol$_{\rm Co}$ was obtained. 
The origin of the higher $\chi_{0}$ for \textit{H}\,$\perp$\,\textit{c} is still unclear. Different background contributions can be excluded.
An effective moment of $\mu_{\mathrm{eff}}$\,=\,3.4\,$\frac{\mu_{\mathrm{B}}}{\mathrm{Co}}$ was determined for both directions. 
Li\textbf{Ca$_2$}[CoN$_2$] follows a Curie-Weiss-law above $\sim$\,120\,K with a ferromagnetic Weiss-temperature of $\Theta_{\mathrm{W}}$\,=\,78\,K and an effective moment of $\mu_{\mathrm{eff}}$\,= 3.2\,$\frac{\mu_{\mathrm{B}}}{\mathrm{Co}}$. 

\begin{figure}
 \begin{center}
  \includegraphics[width=0.6\textwidth]{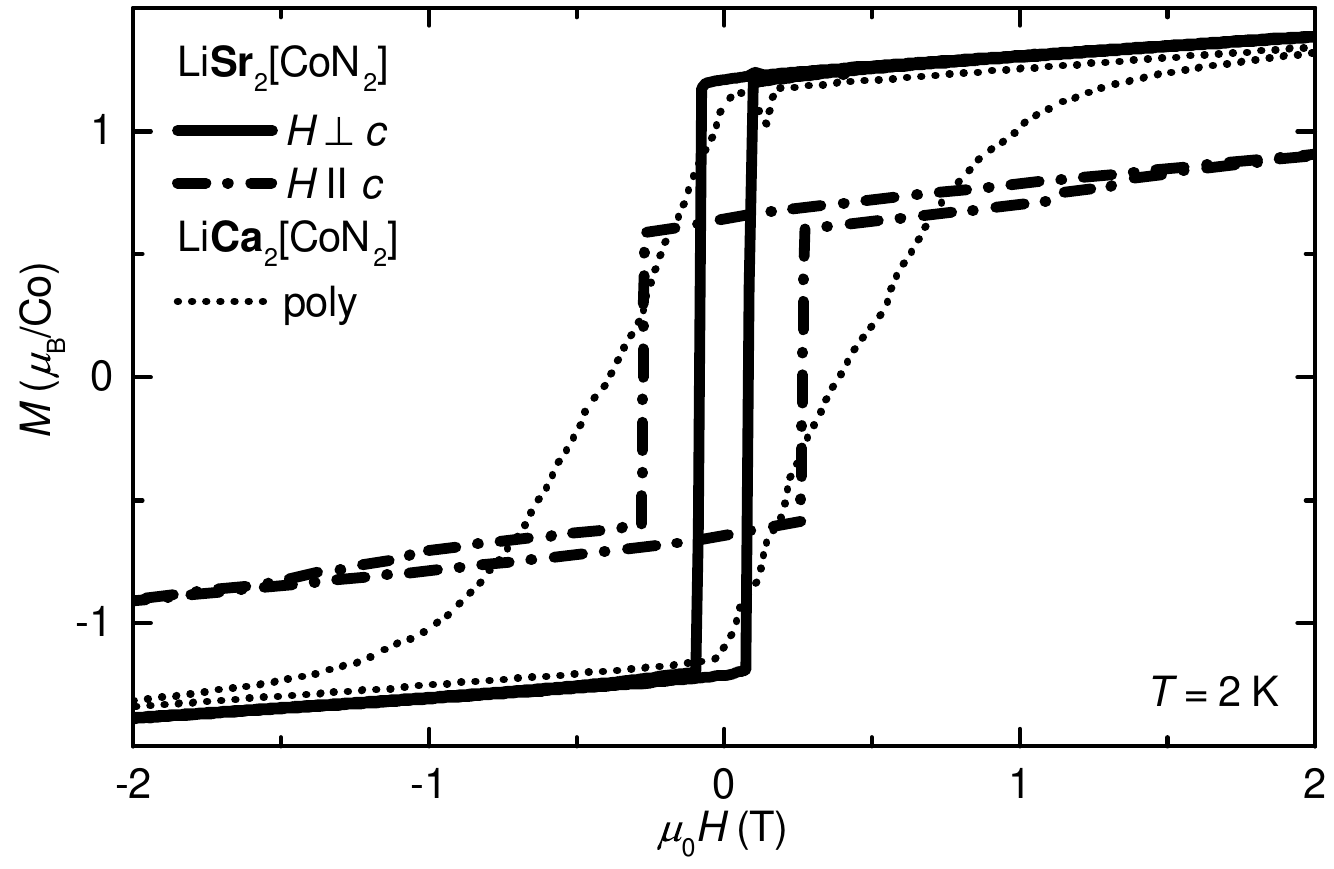}
	\caption{Isothermal magnetization at \textit{T}\,=\,2\,K in $\frac{\mu_{\mathrm{B}}}{\mathrm{Co}}$ as a function of an applied magnetic field. The observed coercivity fields are $\mu_{0}H_{\mathrm{c}}$\,=\,0.1\,T and $\mu_{0}H_{\mathrm{c}}$\,=\,0.3\,T for \textit{H}\,$\perp$\,\textit{c} and \textit{H}\,$\parallel$\,\textit{c}, respectively, for Li\textbf{Sr$_2$}[CoN$_2$] and $\mu_{0}H_{\mathrm{c}}$\,=\,0.4\,T for Li\textbf{Ca$_2$}[CoN$_2$].}
	\label{MvH Vgl}
 \end{center}
\end{figure}

The isothermal magnetization at $T = 2$\,K in Bohr magneton per Co is shown in Fig.\,\ref{MvH Vgl}. 
%Li$_{0.8}$Co\textbf{Sr}$_{2}$N$_{2}$ with \textit{H}$\perp$c and \textit{H}$\parallel$c are represented by the solid and dashed lines, respectivly, Li$_{1.1}$Co$_{0.9}$\textbf{Ca}$_{2}$N$_{2}$ by a dotted line. 
Hysteresis emerges below the transition temperatures for both materials, in accordance with the ferromagnetic ordering inferred from $\chi(T)$. 
In Li\textbf{Sr$_2$}[CoN$_2$] a clear magnetic anisotropy is observed. 
Even in the vicinity of the largest available field of $\mu_{0}H = 7$\,T, \textit{M}(\textit{H}) increases in a linear fashion.
The coercivity fields at $T = 2$\,K amount to $\mu_{0}H_{\mathrm{c}}$\,=\,0.1\,T and $\mu_{0}H_{\mathrm{c}}$\,=\,0.3\,T for \textit{H}\,$\perp$\,\textit{c} and \textit{H}\,$\parallel$\,\textit{c}, respectively. 
Both values are smaller than the coercivity field of $\mu_{0}H_{\mathrm{c}}$\,=\,0.4\,T measured in Li\textbf{Ca$_2$}[CoN$_2$]. 
The changes in $M-H$ observed for Li\textbf{Ca$_2$}[CoN$_2$] are less steep and indicate a superposition of randomly oriented crystallites and domains, with an even higher coercivity field along the magnetically hard axis.

\section{Specific heat}
\begin{figure}
 \begin{center}
  \includegraphics[width=0.6\textwidth]{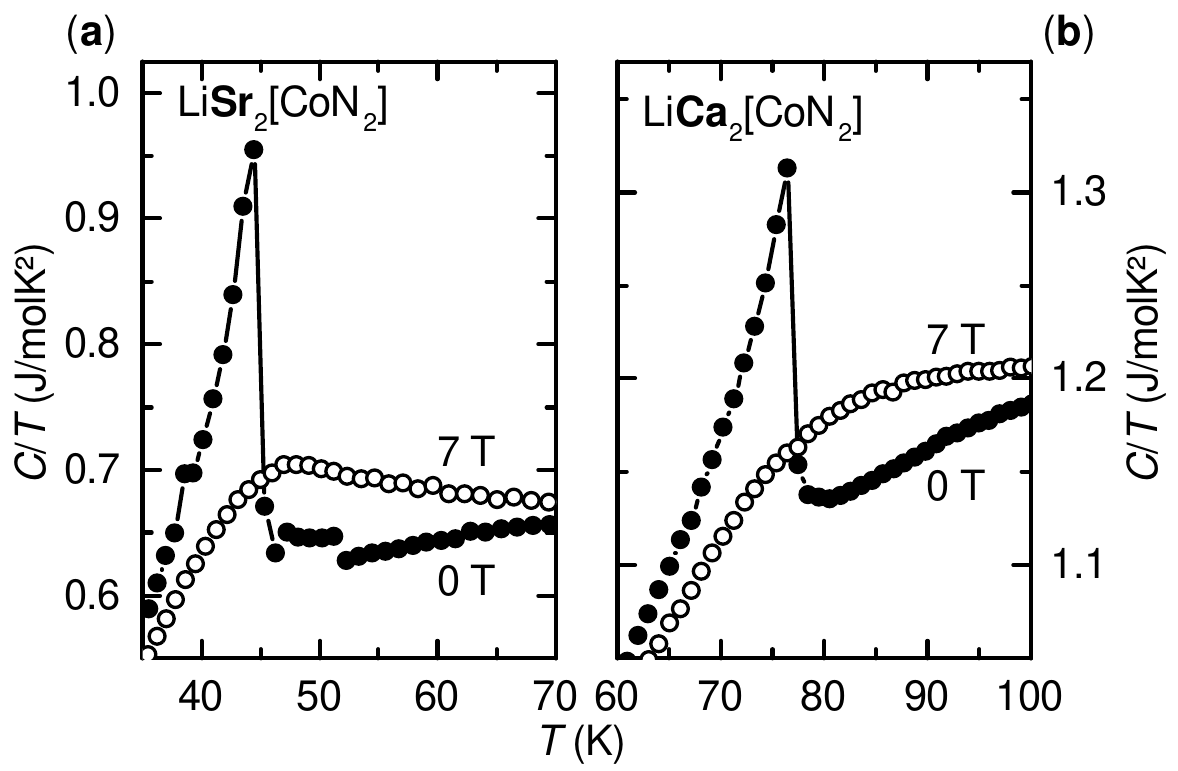}
	\caption{Specific heat of Li\textit{AE}$_2$[CoN$_2$] without (filled circles) and with an applied magnetic field of $\mu_0H = 7$\,T (open circles). (\textbf{a}) \textit{AE}\,=\,Sr. The transition temperature is $T_{\mathrm{C}}$\,=\,45\,K. (\textbf{b}) \textit{AE}\,=\,Ca. The transition occurs at $T_{\mathrm{C}}$\,=\,77\,K. 
In both materials the peak is suppressed in an applied magnetic field of $\mu_0H = 7$\,T.}
	\label{CvT}
 \end{center}
\end{figure}
The specific heat in the region of the phase transition is shown in Fig.\,\ref{CvT}. 
A sharp, $\lambda$-like anomaly is observed at \textit{T}$_{\mathrm{C}}$.
 The transition temperatures of \textit{T}$_{\mathrm{C}}$\,=\,45\,K for Li\textbf{Sr$_2$}[CoN$_2$] and \textit{T}$_{\mathrm{C}}$\,=\,77\,K for Li\textbf{Ca$_2$}[CoN$_2$] are identical to the ones estimated from the magnetic susceptibility. 
The entropy amounts to 1.3\,$\frac{\mathrm{J}}{\mathrm{mol K}}$ for Li\textbf{Sr$_2$}[CoN$_2$] and 0.9\,$\frac{\mathrm{J}}{\mathrm{mol K}}$ for Li\textbf{Ca$_2$}[CoN$_2$] at \textit{T}$_{\mathrm{C}}$. 
In an applied field of $\mu_0H = 7$\,T the anomaly is suppressed and entropy is shifted to higher temperatures.
The measured molar entropy is significantly below the value $R$ln2 = 5.76\,$\frac{\mathrm{J}}{\mathrm{mol K}}$ that is expected for the ordering of a two-level system. 
Extracting the magnetic contribution, as performed for the related system Li$_2$(Li$_{0.7}$Fe$_{0.3}$)N\,\cite{Fix2018b}, is not possible since the non-magnetic reference compound Li\textbf{Sr}$_{2}$[LiN$_{2}$] is structurally unstable. 
Accordingly, the origin for the missing entropy is unclear. 
However, the specific heat measurements were reproduced by using additional samples.
The results revealed an identical behavior with the same transition temperatures and entropies and furthermore proved well defined Co concentrations by using an isotropic, bulk measurement.
%For Li$_{2-x}$Co$_{x}$\textbf{Sr}$_{2}$N$_{2}$ one other single crystal with the same \textit{x}\,=\,1.0 was measured, revealing similar results.
%Further measurements were performed on two more samples of polycristalline Li$_{2-x}$Co$_{x}$\textbf{Ca}$_{2}$N$_{2}$ with similar values of \textit{x}\,=\,0.96 and \textit{x}\,=\,0.94, respectively. They show almost exactly the same behaviour, as described above. 

\section{Quantum chemistry calculations}
The Co $d$-shell electronic structure was analyzed on the theoretical side 
by {\it ab initio} many-body quantum chemistry calculations. For this purpose, we used an embedded cluster defined by the N-Co-N kernel and the adjacent Li and alkaline-earth cations:
[CoN$_2$Li$_{2}AE_8$]$^{13+}$ ($AE\!=\!$ Sr, Ca). For convenience, in our finite-cluster calculations and subsequent analysis of the \textit{ab initio} results, the N-Co-N bonds are taken along the \textit{z} axis.
Due to the rectangular arrangement of Sr (Ca) cations around the N-Co-N link,
the actual point-group symmetry is reduced from $D_{\infty h}$ to  $D_{2h}$. 

As starting point for our computational investigation, complete-active-space self-consistent-field (CASSCF) \cite{Helgaker2000} calculations were performed.
We utilized an active space defined by the five 3$d$ orbitals at the Co site and eight electrons associated with the Co$^{1+}$ $3d^8$ valence configuration; 
to achieve a good description of the lower-energy part of the $d$-$d$ excitation spectrum, 
the orbitals were optimized for an average of the lowest three triplet states. In the subsequent multireference configuration-interaction (MRCI) \cite{Helgaker2000} treatment, 
the N $2s$, $2p$ and Co $3s$, $3p$, $3d$ electrons were correlated.
Spin-orbit couplings (SOC's) were then accounted for according to the procedure described in Ref. \cite{Berning2000}. The solid-state surroundings were modelled as a large array of point charges fitted to reproduce the crystal Madelung field in the cluster region \cite{Klintenberg2000}.

%%%%%%%%%%%%%
%% TABLE 4 %%
%%%%%%%%%%%%%

\begin{table}[h]
\begin{center}
\caption{
Co$^{1+}$ $3d$-shell electronic structure in Li\textit{AE}$_2$[CoN$_2$], as obtained by MRCI calculations. 
Three groups of excited states are found by spin-orbit MRCI, at 7 meV, 250\,-\,450 meV, and 775\,-\,880 meV. 
Other excited states are found above 1.95 eV (not shown in the table). ``  * '' denotes doubly-degenerate levels.
}

\begin{tabular}{llcc|cc|clc}
\hline
\hline\\[-0.10cm]

& $\ $  &$^3\!B_{1g}$   &      &$a\,^3\!B_{2g}$, $a\,^3\!B_{3g}$   &  &$b\,^3\!B_{2g}$, $b\,^3\!B_{3g}$  \tabularnewline%%($\times\!2$)\\
 \hline\\[-0.10cm]
 \textit{AE} = Sr &$\ $MRCI  &0 & &340* &  &825\\
&$\ $  & & & & &830  \\[1cm]
& $\ $MRCI+SOC  &0 & &250* & &810* \\
& $\ $  &7* & &355* & &835 \\
& $\ $  & & &450* & &840 \\
& $\ $  & & & & &870  \\
& $\ $  & & & & &880 \\
 \hline
\textit{AE} = Ca & $\ $MRCI  &0 & &320* & &775\\
&$\ $  & & & & &840 \\[1cm] 
& $\ $MRCI+SOC  &0 & &235* &  &775* \\
& $\ $  &7* & &335* & &785 \\
& $\ $  & & &430* & &850 \\
& $\ $  & & & &  &865 \\
& $\ $  & & & & &875 \\
 \hline
\hline
\end{tabular}
\label{tabLiCoSr2N2}
\end{center}
\end{table}

We found that the low-lying ($<$\,1\,eV) Co$^{1+}$ $d^8$ states are mainly related to the $a_{1g}^2 e_{2g}^4 e_{1g}^2$ and $a_{1g}^2 e_{2g}^3 e_{1g}^3$ configurations, 
with a fully filled apical $d_{z^2}$ orbital.
Whereas the actual symmetry of the cluster is $D_{2h}$, we here employ, for indicating orbital occupations more compact, notations corresponding to $D_{\infty h}$ point-group symmetry; 
the latter describes the symmetry of an isolated N-Co-N kernel.  
In $D_{\infty h}$, $d_{z^2}$ belongs to the $A_{1g}$ irreducible representation, $\{d_{xy},d_{x^2-y^2}\}$ to
$E_{2g}$, and $\{d_{yz},d_{zx}\}$ to $E_{1g}$.
By reducing the symmetry to $D_{2h}$, the $D_{\infty h}$ $E_{2g}$ components change to $A_{g}$ ($d_{x^2-y^2}$) and $B_{1g}$ ($d_{xy}$)  
while the  $E_{1g}$ terms to  $B_{2g}$ ($d_{xz}$) and  $B_{3g}$ ($d_{yz}$). 
The low-energy multiplet structure is then described in $D_{2h}$ by the $^3\!B_{1g}$ triplet ground-state and two pairs of $^3\!B_{2g}$, $^3\!B_{3g}$ terms.

MRCI relative energies describing the excitation spectrum of a Co$^{1+}$ ion in Li\textbf{Sr$_2$}[CoN$_2$] are listed in Table \ref{tabLiCoSr2N2} (\textit{a} and \textit{b} indices are used to distinguish between states of the same symmetry and spin multiplicity in the calculations without SOC).
The spin-orbit treatment was carried out in terms of all high-spin ($S\!=\!1$) states and those low-spin ($S\!=\!0$) components with MRCI relative energies up to 3.5 eV.
Three sets of low-energy spin-orbit states are found in the calculations: 
low-lying spin-orbit states arising from the $a_{1g}^2 e_{2g}^4 e_{1g}^2$
configuration below 10 meV plus other states mainly arising from the $a_{1g}^2 e_{2g}^3 e_{1g}^3$ configuration at 250\,-\,450 and 810\,-\,880 meV (see Table \ref{tabLiCoSr2N2}). 
Other excited states lie above 1.95 eV by MRCI and are not listed in the table.

Test calculations in which only the lowest $^3\!B_{1g}$ term is included in the spin-orbit treatment yield degenerate spin-orbit states,
confirming that the $a_{1g}^2 e_{2g}^4 e_{1g}^2$ configuration [$d_{z^2}^2 (d_{x^2-y^2} d_{xy})^{4} (d_{xz} d_{yz})^2$] is associated with second-order SOC's \cite{Bryan2012}. In other words, the finite splitting among the $^3\!B_{1g}$ levels is the result of interactions with higher-lying states, not an intrinsic feature of the [$d_{z^2}^2 (d_{x^2-y^2} d_{xy})^{4} (d_{xz} d_{yz})^2$] configuration. 

Similar type of results are provided for the Li\textbf{Ca$_2$}[CoN$_2$] compound in Table \ref{tabLiCoSr2N2}. The main difference as compared to Li\textbf{Sr$_2$}[CoN$_2$] is having larger splittings within the group of $b\,^3\!B_{2g}$ and $b\,^3\!B_{3g}$ (related) states.

\section{Discussion}

Characteristic values associated with the occurrence of magnetic ordering are summarized in Tab.\,\ref{tabmag}.
The $T_{\rm C}$s obtained by different methods are in good agreement.
For the smaller $AE$ = Ca, the ordering temperature almost doubles.
This is considered a direct consequence of the smaller ionic radius of Ca, that leads to a smaller separation of the one-dimensional N-Co-N chains.
The weaker inter-chain coupling for the larger $AE$ = Sr yields a lower $T_{\rm C}$ in accordance with the Mermin-Wagner theorem\,\cite{Mermin1966}, that forbids the emergence of magnetic ordering in a purely one-dimensional system. 

The effective magnetic moments of roughly $3.3\,\mu_{\rm B}$ per Co are moderately enhanced when compared to the spin-only value of 2.82\,$\mu_{\mathrm{B}}$ and indicate an orbital moment contribution.
Similar values were observed in Co-doped Li$_{3}$N \cite{Jesche2015}, which shows the same coordination. 
However, no magnetic hysteresis was observed for the latter. 

The coercivity fields of $\mu_0H_c = 0.3$\,T and $\mu_0H_c = 0.4$\,T for $AE$ = Sr and Ca (Fig.\,\ref{MvH Vgl}), respectively, are considered large for a 3\textit{d} transition-metal based material. 
Similar or even larger values are only found in nanostructured systems: e.g. Co nanotubes with $\mu_0H_c = 0.13$\,T at $T = 5$\,K\, \cite{Graf2006}, surface particles in cold pressed LaCoO$_3$ with $\mu_0H_c = 1$\,T at $T = 5$\,K\,\cite{Yan2004}, and Co nanowires with $\mu_0H_c = 1.1$\,T at RT\,\cite{Gandha2014}. 

\begin{table}
  \begin{center}
   \caption{Characteristic values associated with magnetic order in Li\textbf{Sr$_2$}[CoN$_2$] and Li\textbf{Ca$_2$}[CoN$_2$]. Shown are the values of \textit{T}$_{\mathrm{C}}$ determined from magnetic susceptibility ($\chi$) and specific heat (HC) as well as the inferred Weiss-temperatures ($\Theta_{\mathrm{W}}$), effective magnetic moments ($\mu_{\mathrm{eff}}$), coercivity fields at $T = 2$\,K ($\mu_{0}H_{\mathrm{c}}$) and entropy (\textit{S}).}
   \begin{tabular}{lccc}
	\hline 
	\hline
  & \textit{AE}\,=\,Sr & & \textit{AE}\,=\,Ca\\
 &  \textit{H}\,$\parallel$\,\textit{c} & \textit{H}\,$\perp$\,\textit{c}   &  poly\\ 
  \hline
  \textit{T}$_{\mathrm{C}}$($\chi$)  & 44\,K & 44\,K & 76\,K\\
  \textit{T}$_{\mathrm{C}}$(HC)  & 45\,K & 45\,K & 77\,K\\
  $\Theta_{\mathrm{W}}$ & 42\,K & 46\,K  & 78\,K\\
  $\mu_{\mathrm{eff}}$  & 3.3\,$\frac{\mu_{\mathrm{B}}}{\mathrm{Co}}$ & 3.3\,$\frac{\mu_{\mathrm{B}}}{\mathrm{Co}}$ & 3.2\,$\frac{\mu_{\mathrm{B}}}{\mathrm{Co}}$\\
  $\mu_{0}H_{\mathrm{c}}$ & 0.3\,T & 0.1\,T & 0.4\,T \\
  \textit{S}(\textit{T}$_{\mathrm{C}}$)  & 1.3\,$\frac{\mathrm{J}}{\mathrm{molK}}$ & 1.3\,$\frac{\mathrm{J}}{\mathrm{molK}}$ & 0.9\,$\frac{\mathrm{J}}{\mathrm{molK}}$\\ 
  
  \hline 
  \hline
  \end{tabular}
 \label{tabmag}
 \end{center}
\end{table}

%* $\mu_{0}$\textit{H}\,=\,0\,T
%For Fe-doped Li$_{3}$N it amounts to 6.5\,$\frac{\mu_{\mathrm{B}}}{\mathrm{Fe}}$ and is almost twice as much, as the one observed in the title compound. The effective moment of Fe-doped Li$_{4}$SrN$_{2}$ amounts to 5.4\,$\frac{\mu_{\mathrm{B}}}{\mathrm{Fe}}$ and is significantly larger than the Co-substituted compounds \cite{Jesche2014b, Hohn2016}.

Li\textbf{Sr$_2$}[CoN$_2$] reveals the unusual feature of the remnant moment being smaller in the direction of the broader hysteresis.
Similar behavior has been observed in CeRuPO\,\cite{Krellner2008} and CeFeAs$_{1-x}$P$_x$O\,\cite{Jesche2012b} and is attributed to a competition between anisotropic exchange interactions and single-ion anisotropy. 

The measured magnetization at $T = 2$\,K does not reach the spin-only value of $2\,\mu_{B}$ per Co for $AE$ = Sr and Ca (in contrast to the effective moments deduced from the high temperature part of the susceptibility).  
This is caused by the mutually perpendicular orientation of the different N-Co-N 'molecule axes', which run along $[1\,1\,0]$ and $[1\,\bar{1}\,0]$, and determine the orientation of the easy axes.
The maximum applied field of $\mu_0H = 7$\,T is not sufficient to overcome the in-plane anisotropy and the maximum projection of the ordered moment amounts to $\mu_{\rm sat}/ \sqrt{2}$ (along $<$1\,\,0\,0$>$).
Furthermore, the large remnant magnetization for \textit{H}\,$\parallel$\,\textit{c} indicates a tilt of the spontaneous moment away from the $ab$-plane or, in other words, a sizable projection of the spontaneous moment towards the $c$-axis. 

Since the magnetically easy axis is determined by the transition-metal valence state and the orientation of the 'molecule axis'\,\cite{Jesche2015, Ke2015}, a possible explanation for the canted spontaneous moment is given by an intermediate or mixed Co valence caused by the Li vacancies mentioned earlier (see Sec. \ref{Crystal growth}).
Co(II) would correspond to a $3d^7$ electron configuration as proposed for Fe-doped Li$_{3}$N\,\cite{Jesche2014b} and Fe-doped Li$_{4}$SrN$_{2}$\,\cite{Hohn2016}, which do show the easy axis parallel to the molecule axis. 
Co(I), on the other hand, corresponds to $3d^8$ and easy axis perpendicular to the 'molecule axis'\,\cite{Jesche2015}.
With the data at hand, we cannot completely rule out the presence of two Co species. 
Even an itinerant scenario could account for the observed behavior: the magnetic anisotropy changes sign for an average electron count of $3d^{7.8}$ Ref.\,\onlinecite{Jesche2015}. 
This is consistent with the measured Li vacancy concentration of 0.2 per formula unit.

%Tha saturation moment is almost the same for Li$_{0.8}$Co\textbf{Sr}$_{2}$N$_{2}$ and Li$_{1.1}$Co$_{0.9}$\textbf{Ca}$_{2}$N$_{2}$, if field is applied parallel to the effective linear molecule of Li$_{0.8}$Co\textbf{Sr}$_{2}$N$_{2}$. 
%For Li$_{1.1}$Co$_{0.9}$\textbf{Ca}$_{2}$N$_{2}$ the direction of the field is unknown. Here $H_{\mathrm{c}}$ is slightly higher, than it is for Li$_{1.1}$Co$_{0.9}$\textbf{Sr}$_{2}$N$_{2}$ and the available data suggests, that even higher values might be revealed, if measurements on single crystals could be made. Substitution of Ca with Sr significantly lowers $T_{\mathrm{C}}$, but $\mu_{\mathrm{eff}}$ remains nearly the same.

An inherent correlation seems to exist between the single-ion anisotropy and the presence of spontaneous magnetization in linearly coordinated materials:
there is no hysteresis/remnant magnetization observed in the easy plane Mn- and Co-doped Li$_3$N systems\,\cite{Jesche2015}, which show easy axis perpendicular to the 'molecule axis' (N-Mn-N and N-Co-N, respectively).
Fe-based compounds show the easy axis along the 'molecule axis' (Fe-doped Li$_3$N\,\cite{Jesche2014b} and Li$_4$SrN$_2$\,\cite{Hohn2016}) and also exhibit spontaneous magnetization with comparatively large coercivity fields. However, an easy axis parallel to the molecule axis is not sufficient for the emergence of hysteresis as shown by Ni-doped Li$_{3}$N \cite{Jesche2015}.

The similarities of the magnetic properties of Fe-doped Li$_3$N and the title compound hold even though the nature of the spontaneous magnetization differs: 
Li\textbf{Sr$_2$}[CoN$_2$] shows a well defined phase transition into the ordered state as evidenced by the sharp anomaly in specific heat (Fig.\,\ref{CvT}).
In Li$_2$(Li$_{0.7}$Fe$_{0.3}$)N, on the other hand, a crossover from high temperature paramagnetic to a slowly relaxing, seemingly static low temperature state takes place that is driven by an enhancement of the spin relaxation times with decreasing temperature. 
The corresponding specific heat shows a broad Schottky anomaly\,\cite{Fix2018b}. 
The different behavior is not caused by the larger dilution of Fe in Li$_2$(Li$_{0.7}$Fe$_{0.3}$)N, which corresponds to a total concentration of 7.5\% Fe per formula unit and an average Fe-Fe distance of 5.4\,\AA. 
The full Co occupation in Li\textbf{Sr$_2$}[CoN$_2$] leads to similar average Co-Co distances of 5.8\,\AA, however, the substitution takes place in an ordered fashion.
Accordingly, the exchange interaction in the two materials is substantially different whereas the local moment properties lead to comparable magnetic anisotropies. 

%The coercivity field of Co-doped Li$_{3}$N is 0, but for Fe-doped Li$_{3}$N it amounts to $\mu_{0}H_{\mathrm{c}}$\,=\,11\,T! and is significantly larger, than the one observed in the title compound. Perpendicular to the effective linear molecule, the hysteresis is small, or vanishes completely, depending on the Fe-concentration, while it is present in Li$_{0.8}$Co\textbf{Sr}$_{2}$N$_{2}$, independent of the direction of the applied magnetic field. However measurements were performed for one Co-concentration only. The dependence of the hysteresis on the Co-concentration would be subject to further research. 

%Fe-doped Li$_{4}$SrN$_{2}$ shows behaviour similar to Fe-doped Li$_{3}$N. Hysteresis can only be observed if field is applied parallel to the effective linear molecule. The coercivity field of $\mu_{0}H_{\mathrm{c}}$\,=\,7\,T is smaller than the one observed in Fe-doped Li$_{3}$N, but still higher than the coecivity fields found in Li$_{2-x}$Co$_{x}$\textit{AE}$_{2}$N$_{2}$. However $H_{\mathrm{c}}$ is still high compared to typical rare-earth-free ferromagnets.

\section{Summary}
Li\textbf{Sr$_2$}[CoN$_2$] single crystals up to several millimeter along a side were obtained from a Li-rich flux.   
Small single crystals sufficient for X-Ray diffraction and good quality polycrystalline samples were obtained for the isostructural Li\textbf{Ca$_2$}[CoN$_2$] compound.
The title compound orders ferromagnetically with \textit{T}$_{\mathrm{C}}$\,=\,45\,K whereas a significantly enhanced \textit{T}$_{\mathrm{C}}$\,=\,78\,K was found for the Ca analog. The magnetic anisotropy, coercivity, and effective moment are moderately enhanced and indicate the presence of an orbital contribution to the magnetic moment and large single ion anisotropies. 
Although this occurs through second-order spin-orbit couplings, the resulting magnetic anisotropy energy is still large, in the region of 7\,meV as shown by quantum chemistry calculations.
Thus we have shown another example of a linearly coordinated transition metal, different from Fe, that exhibits orbital moment contributions and comparatively large magnetic anisotropies.

\section*{Acknowledgments}
The authors gratefully acknowledge L. K\"ohler and J. Richter for experimental help and S. H\"uckmann and Y. Prots for X-ray powder and single crystal diffraction data collection. A. Mohs, A. Herrnberger and K. Wiedenmann are acknowleged for technical support. Financial support was provided by the Deutsche Forschungsgesellschaft (DFG, German Research Foundation) - grants No. JE 748/1 and HO 4427/3.

\end{document}